%% file: main.tex
\documentclass[twocolumn]{article}
\usepackage{amsmath,epsfig,latexsym,psfig,url,graphicx}
\usepackage[top=.9in,bottom=.1in,left=.9in,right=.9in]{geometry} 

\renewcommand{\footnotesize}{\small}

\def\up{\vspace*{-1cm}}

\def\Section {\S}

\def\compactify{\itemsep=0pt \topsep=0pt \partopsep=0pt \parsep=0pt}
 \let\latexusecounter=\usecounter

\title{\up Markets are Dead, Long Live Markets}

\newcount\aucount
\newcount\originalaucount
\newdimen\auwidth
\newdimen\auskip
\newcount\auskipcount
\newdimen\auskip
\global\auskip=1pc
\newdimen\allauboxes
\allauboxes=\auwidth
\newtoks\addauthors
\newcount\addauflag
\global\addauflag=0 

\gdef\numberofauthors#1{\global\aucount=#1
\ifnum\aucount>4\global\originalaucount=\aucount \global\aucount=4\fi 
\global\auskipcount=\aucount\global\advance\auskipcount by 1
\global\multiply\auskipcount by 2
\global\multiply\auskip by \auskipcount
\global\advance\auwidth by -\auskip
\global\divide\auwidth by \aucount}

\newfont{\affname}{phvr at 10pt}
\newfont{\eaddrfnt}{phvr at 8pt}
\newfont{\authorfont}{phvr at 12pt}

\def\alignauthor{
\end{tabular}%
  \begin{tabular}[t]{p{\auwidth}}\centering}%

\numberofauthors{4}

\author{
      \alignauthor \authorfont{Kevin Lai}\thanks{kevin.lai@hp.com, Information Dynamics Laboratory, HP Labs, Palo Alto, CA 94304}
          }

\date{}

\begin{document}

\maketitle

\input{abstract}

\input{introduction}

\input{example}

\input{economics}

\input{simulation}

\input{integration}

{\footnotesize
\bibliographystyle{abbrv}
\bibliography{../bibliographies/resource_allocation,../bibliographies/economics,../bibliographies/networking,../bibliographies/overlay,../bibliographies/network_performance,../bibliographies/peer-to-peer,../bibliographies/security,../bibliographies/reputation,../bibliographies/network_architecture,../bibliographies/grid,../bibliographies/scheduling,../bibliographies/virtualization,../bibliographies/game_theory,../bibliographies/algorithms,../bibliographies/transport_protocols}
}

\end{document}

%% file: abstract.tex
\begin{abstract}
\small
Researchers have long proposed using economic approaches to resource
allocation in computer systems. However, few of these proposals became
operational, let alone commercial. Questions persist about the
economic approach regarding its assumptions, value, applicability, and
relevance to system design. The goal of this paper is to answer these
questions. We find that market-based resource allocation \emph{is}
useful, and more importantly, that mechanism design and system design
should be integrated to produce systems that are both economically and
computationally efficient.

\end{abstract}

%% file: introduction.tex
\section{Introduction}
\label{sec:introduction}
A key advantage of the Internet, peer-to-peer file sharing networks,
and systems like PlanetLab \cite{planetlab2003} is the sharing of
computational resources. This provides a variety of benefits,
including higher utilization, increased throughput, lower delay (due
to dispersion of resources in the network), and higher reliability.
However, resource allocation remains an issue. The problem is how to
allocate a shared resource fairly, with economic efficiency (where
efficiency is the ratio of the total actual benefit to all users to
the optimal benefit), and at low cost.

Scheduling algorithms like Proportional Share are a partial
solution. The problem with PS is determining how to set the
weights. Assuming that the values of tasks varies over time, no single
set of weights will suffice. Setting of weights cannot be left to
users because they have a strong incentive to always ask for the
highest possible weight. Having the system administrator set weights
is error-prone and time-consuming. 

Economics and game theory offer an alternative. The area of
\emph{mechanism design} is concerned with algorithms where individuals
optimizing their own utility results in high overall utility. A market
(or auction) is an example. In the resource allocation context, as users
optimize the benefit that they receive from their applications, the
mechanism optimizes the overall efficiency of the resource allocation,
without the intervention of an administrator.

This is not a novel idea. Researchers \cite{ferguson1988}
\cite{malone1988} proposed this approach as early as 1988, and there
were likely earlier ones. Since then, several researchers
\cite{waldspurger1992} \cite{regev1998} \cite{stratford1999}
\cite{chun2000} \cite{ng2003}
have pursued it. Unfortunately, there have been few implementations
\cite{waldspurger1992} \cite{chun2000} \cite{coleman2004}.  Given $17$
years of research, it is surprising that there is not even one
commercially or freely available system for market-based resource
allocation. Given the impact of poor resource allocation on systems
like PlanetLab, this appealing approach and a large body of
enthusiastic publications, why have so few systems been built?  We
believe the lack of operational systems is because questions persist
in the minds of system designers about the value of the economic
approach, its applicability, and its relevance to system design.

In this paper, we examine some of these questions using a combination
of qualitative arguments and simulation. We do not claim to have
conclusive answer. Instead, we hope to provide sufficient affirmative
evidence that more real systems should be built, deployed, and
evaluated.

%% file: example.tex
\section{An Example}
\label{sec:example}

In this section, we examine Proportional Share (PS) scheduling as an
example of the problem that market-based resource allocation seeks to
solve. PS gives user $i$ with weight $w_i$ a $w_i / \sum w$ share of
the resources. Using PS hierarchically, the user can assign these
resources to his tasks. For example, if Alice has weight $2$ and Bob
has a weight of $1$, then Alice has two-thirds of the total resources
for her tasks and Bob has one third. Suppose that this is necessary
because Alice must process twice as many queries as Bob. This works
well if Alice is always doing something more important than Bob, but
sometimes Bob will have an important task (e.g., serving a client
query) while Alice has a much less important task (e.g.,
non-time-critical background jobs like garbage collection). In this
case, Alice's task will still get most of the resources. This is an
economically inefficient situation because although Alice receives
some small benefit, Bob receives much less than he could, and the
total benefit is much less than if Bob received most of the current
resources.  This is a common situation because for most users and
their applications, the arrival process of important work is highly
bursty (e.g., a web/email/file server).

One solution is to rely on Alice to yield resources (e.g., using
\texttt{nice} or other means to set lower weights on tasks) when doing
less important tasks. Unfortunately, Alice has an incentive to deny
Bob the resources and use them herself.  This is an example ``Tragedy
of the Commons''~\cite{hardin1968} where optimizing for individual
utility results in low overall utility. On the other hand, optimizing
for overall utility depends on knowing the relative values of every
task being run. Unfortunately, users cannot be relied on to accurately
report these values without an honesty incentive. Some users may
behave \emph{obediently} by yielding resources or honestly reporting
task values, but those that do not (\emph{strategic} users) lower the
efficiency of the system, and, worse, provide an incentive for
obedient users to become strategic.

Another solution is to have a system administrator monitor the system
and dynamically change the weights of the users to maintain high
efficiency. However, this is expensive, time-consuming, and
error-prone, and it is does not scale to large numbers of users and
resources. 

Market-based resource allocation addresses this problem. For example,
Alice and Bob are issued a currency with income rates in the ratio $2$
to $1$. They use these credits to bid for resources in a market where
user $i$ with bid $b_i$ receives a $w_i / \sum w$ share of the
resources.  When a user has used her share, then her bid is deducted
from her balance of credits. Since Alice is aware that garbage
collection is a less critical task than serving client queries, she
will spend fewer credits when doing the former and more doing the
latter. The mechanism provides an incentive for users to truthfully
reveal how much they value resources. This allows Bob to get more
resources than Alice when he is processing queries and she is doing
garbage collection. Over the long-term, Alice can still process twice
as many queries as Bob (assuming similar workloads).

%% file: economics.tex
\section{Resource Allocation Markets}
\label{sec:economics}

In this section, we examine some questions about market-based resource
allocation.

\subsection{What benefits do markets provide over long-term PS?} 
\label{sec:fixed_pricing}

In long-term Proportional Share, user $i$'s share of the resources is
$w_i / \sum w$ over a longer time period (e.g., a week or year) than
the $10$ milliseconds of a typical CPU scheduler. This provides some
of the flexibility of a market-based system. 

However, long-term PS is not sufficient to reach economic
efficiency. It does not encourage users to shift usage from high
demand periods to low demand periods. It also does not encourage users
to shift usage from high demand resources to low demand resources. For
example, a system could have high demand for CPU cycles, but low
demand for physical memory. In a typical application of PS, the CPU
and memory would be allocated separately and applications would have
no incentive to use more memory and fewer CPU cycles \cite{sullivan2000}. 

\begin{figure}[htb]
\includegraphics[width=8cm]{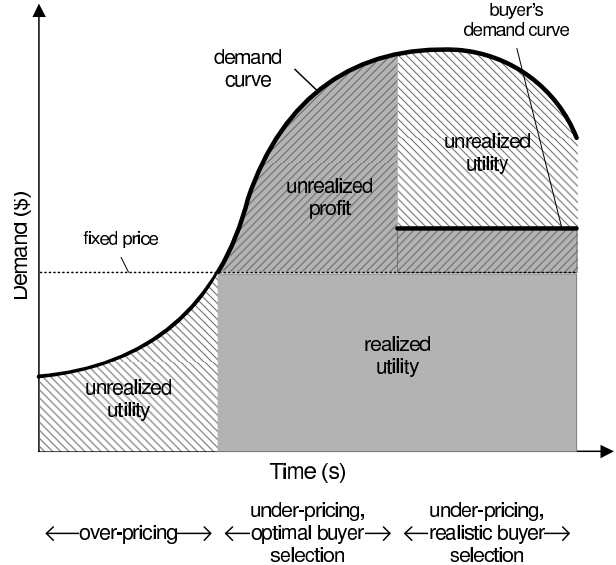}
\caption{\small {\bf Fixed and Variable Pricing.} This figure shows a
  variable demand curve over time and how its efficiency compares to a
  fixed pricing curve.}
\label{fig:pricing}
\end{figure}

A related solution is to use fixed pricing. For example, each CPU
cycle costs \$10 and each memory page costs \$1. A user in this
example has a strong incentive to use more memory to save CPU cycles,
so this begins to address the multi-resource problem. Pricing also
provides a way for applications to express quality-of-service needs.
For example, an application may not need many cycles, but instead
needs them quickly after an interrupt has occurred
\cite{duda1999}. Time slices could be priced differently based on how
quickly they are scheduled.

The problem is how these prices should be
set. Figure~\ref{fig:pricing} shows that no fixed price is as
efficient as a variable price, assuming variable demand.  In the left
regime, the demand is below the fixed price, so the buyer is unwilling
to buy the resource and the utility that the buyer would have gained
by using the resource is unrealized (indicated by the striped area
under the demand curve). In the middle regime, someone is willing the
pay the fixed price, so that buyer is able to use the resource and
gains some utility (indicated by the gray area under the demand
curve). Assuming that the seller chooses the optimally efficient buyer
(i.e., the one willing to pay the most), then the difference between
the demand curve and the fixed price is unrealized profit for the
seller (indicated by the striped gray area).  Unrealized profit
contributes to overall inefficiency in some cases (see
\Section~\ref{sec:money}). However, without a covert channel, a fixed
price seller cannot distinguish among the demand curves of potential
buyers, so the actual buyer's demand curve will probably be lower than
the optimal buyer's and the area between them is more unrealized
utility. In general, the more variable the demand, the worse the
efficiency loss of fixed prices.

One definition of a market is a way to set prices so that they follow
the demand. As a result, most, if not all of the utility under the
demand curve in Figure~\ref{fig:pricing} would be realized. 

\subsection{Are markets fair?}
\label{sec:fairness}

Markets allow users to save currency. Could this allow a user to save
enough currency to starve out other users? Are markets unfair in some
other way?

Whether markets are fair depends on the market and the definition of
fair. One definition is that all users receive resources in proportion
to an exogenously determined weighting system. For example, if Alice
has a weight of $2$ and Bob has a weight of $1$, Alice should get
twice the resources of Bob over an arbitrarily long time interval.
Assuming that Alice and Bob have demand curves that equally correlated
with the overall demand, the market described in
\Section~\ref{sec:example} is fair by this definition. 


One variation of this definition is to restrict the timescale used to
measure the resource usage. For example, within an hour interval, if
Alice and Bob want resources, she should always get twice the
resources that he gets, even though he saved up his credits and she
spent hers. This is useful to prevent starvation by those who
mis-manage their resources. Another possibility is that Bob saves up
credits over a long period of time and then spends them all at once,
thus starving out Alice. In both cases, the system can monitor Alice
and Bob's credits and redistribute credits from the wealthy to the
poor. This reduces the degree of unfairness, but also reduces
efficiency because Alice and Bob have a reduced incentive manage
resource usage carefully.

The conclusion is that markets are not inherently unfair. System
designers can tune a market-based system to make the tradeoff between
efficiency and fairness that is appropriate for their users. 

\subsection{Are markets useful when real money is not involved?}
\label{sec:money}
Economic mechanisms provide an intuitive mapping between resource
allocation and a business model for selling resources. However, in
some cases, the resources are just being shared and not sold (e.g.,
employees sharing machines in their company's data center). 

Markets are still useful in these situation. The simplest
configuration is an open-loop economy, where the resource owner issues
credits to users who can then spend them on resources. The main
efficiency gain results from users having an incentive to truthfully
reveal the value of their tasks (as shown in \Section~\ref{sec:simulation}). 

Another alternative is a closed loop economy where users both consume
and provide resources. PlanetLab could be run this way. A closed loop
economy provides more incentives for efficiency than an open loop
one: as prices rise, so does providers' profit, which increases
the incentive to provide resources. This raises competition and
eventually causes prices to fall. At no point in this cycle is real
money necessarily involved. 

\subsection{Are markets predictable?}
\label{sec:predictability}

Markets may allocate resources efficiently on average, but prices for
resources fluctuate, so how can users predict the cost of the
resources that they need? 

In this context, we define predictability (i.e., performance
isolation or quality-of-service) as the ability to provide a fixed amount
of resources over a period of time with high probability,
regardless of the demand put on the system. An example of an
application needing predictability is a web server that needs to serve
$n$ requests per second with $99\%$ of requests served within $d$
seconds. 

The market from \Section~\ref{sec:example} can provide this capability
by adding the ability to reserve fixed shares, where a share is a
fixed percentage of a resource (e.g., $.1\%$ of a $1$ GHz CPU = $1$
MHz). These shares have a fixed duration and are sold using an
auction. The operator of the example web server calculates the
resources necessary to meet his needs and bids for those
resources. The cost of this approach is that resource may be
under-utilized because some resources may be reserved, but go unused.


Although similar to techniques used in non-market systems
\cite{stoica1997} \cite{sullivan2000} \cite{duda1999}, the market
allows the predictability mechanism to be used more efficiently. The
problem with these systems is the difficulty in deciding how much of
the resource should be devoted to best-effort service and how much to
reserved service. The optimal split will likely vary significantly
over time.  Users would prefer reserved service if the cost to them is
equal, but reserved resource are less efficient than best-effort
because of the potential for under-utilization. The benefit of the
market is in forcing users to consider whether they really require
reserved resources and in helping the system determine
reserved/best-effort split. High bids for reserved service 
will cause users who can tolerate best-effort to do so and indicate to
the system to reserve more resources. Low bids will do the opposite.


%% file: simulation.tex
\section{Simulation Results}
\label{sec:simulation}

In this section, we present preliminary simulation results quantifying
the efficiency gains of a market. The basic idea is to simulate a
single CPU server running the CPU-intensive tasks of several users. We
examine different resource allocation algorithms and different user
behaviors.

\begin{table}
\small
\begin{center}
\begin{tabular}{|c|c|}
\hline
Parameter & Value  \\
\hline
\hline
Users & 10 \\
\hline
Running Time & 1000s \\
\hline
Task Interarrival & Gaussian, $\mu$: [1s, 120s], $\sigma$: $\mu/2$ \\
\hline
Task Size & Gaussian, $\mu$: 10, $\sigma$: 5 \\
\hline
Task Deadline & Gaussian, $\mu$: 75, $\sigma$: 37.5 \\
\hline
Task Value & Uniform, range: (0, 1] \\
\hline
\end{tabular}
\end{center}
\label{tab:parameters}
\caption{Simulation parameters}
\end{table}

The simulation parameters are summarized in
Table~\ref{tab:parameters}.  A user may have more than one pending
task, but users only run one task at a time. If a task completes by
the deadline, then the user receives $value * size$ utility, otherwise
There is one server providing resources for tasks. A task finishes
when it accumulates resources equal to its size. A user can run one
task, switch to a new task, and then switch back to the first task
without cost. 

The server uses one of two resource allocation schemes: Proportional
Share or Market Proportional Share. With Proportional Share, the
server allocates its resources to tasks according to the weight
assigned by the task's owner. With Market Proportional Share, users
have an income of \$$1$ credit per second. If Alice spends $1$ credit
and Bob spends $2$, then Alice's task gets $.66$ resources, while
Bob's task gets $.33$. The income can be saved.

We simulate three different user behaviors: obedient, strategic
without a market, and strategic with a market. Obedient users assign a
weight to their tasks equal to the task's value. Non-market strategic
users assign the maximum possible weight to all of their tasks. Market
strategic users budget their credits according to the task. The idea
is to spend more credits on more valuable tasks and to apportion the
credits over the lifetime of the task. Market strategic users assign
the following credits per second to run their most valuable task:
$(balance * value)/ (deadline - now)$.  $balance$ is the user's
current credit balance, $value$ is the value of the user's most
current valuable task, $deadline$ is the deadline of that task, and
$now$ is the current time.

\begin{figure}[htb]
\hspace*{-.6cm}
\includegraphics[angle=270,width=9cm]{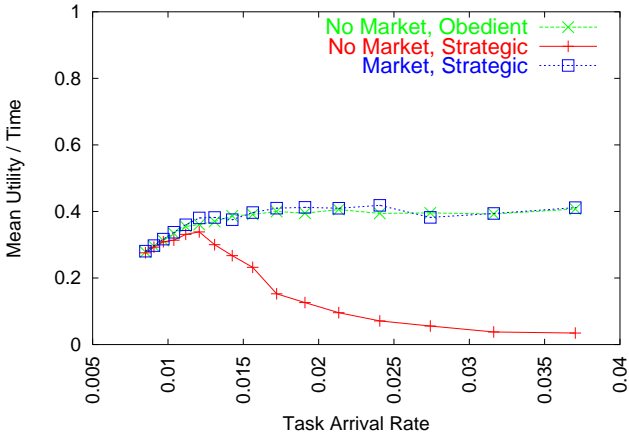}
\caption{\small The utility of different user behaviors 
and mechanism as a function of system load. }
\label{fig:market_no_market}
\end{figure}

Figure~\ref{fig:market_no_market} shows the simulation results. The
y-axis is the mean utility per host per time unit. The x-axis shows
the mean task arrival rate in the system and is a measure of overall
system load. Each point in the graph is a run of the simulator.

As the load increases to the right, the obedient users without a
market are able to maintain a high level of utility. However, users
have no incentive to be unilaterally obedient. Instead, their
incentive is to strategically give a high weight to all their tasks.
The plot of the non-market strategic users shows that they are able to
maintain a high level of utility when the system is lightly loaded
(from 0.0 to 0.0125), but as the load saturates the system, utility
drops to zero. At this point the system wastes resources running tasks
that never meet their deadlines and therefore provide no utility.  In
a system without a mechanism or significant social pressure, some
users inevitably become strategic. To counter this, we use the market
mechanism. The strategic users are forced to truthfully reveal the
value of their tasks and the system can maintain the same high level
of utility as when all users were obedient.


%% file: integration.tex
\section{Integrated Mechanism and System Design}
\label{sec:integration}

Mechanism design is traditionally part of economics while system
design is part of computer science. Why should they be done in
concert? How much benefit would be provided by adding an existing
mechanism, such as bartering or EBay, to a separately designed
system?

A long-standing principle \cite{levin1975} in system design is to
separate policy and the computational mechanism used to implement the
policy (not the economic mechanism in the mechanism design
sense). However, as Clark, et al. \cite{clark2002} point out, policy
and computational mechanism cannot truly be separated because the
mechanism defines what policies are possible. 

This would not be a problem if computational mechanism designers
provided interfaces for efficient and scalable policies, but this has
not been the case. For example, the market from
\Section~\ref{sec:example} requires statistics on resource usage (e.g., CPU
cycles, memory pages, disk blocks) and dynamic control over
allocation. However, many systems do not export detailed information
on usage or allow dynamic control of allocations
\cite{waldspurger2002}. Another example is that several computational
mechanisms assume a bartering policy. However, bartering economies
have very little fluidity. It is difficult to find a mutually
satisfying partner for each transaction and the complexity of
determining the exchange rates of $n$ resources is $O(n^2)$.

Even using an efficient economic mechanism from other contexts can
result in poor efficiency in a computational environment. Unlike many
other resources, the latency to access a computational resource is
critical because changes in demand are unpredictable. For example,
one possible economic mechanism for computational resources is to
auction them on EBay. Auctions on EBay are tuned for human bidding so
they typically take hours to close. The problem is that a web server's
demand may be spiking right now. By the time the auction closes, the
high load will have dissipated. The web server's operator could try to
anticipate load and purchase capacity in advance, but this results in
unused capacity.

In general, a pure mechanism designer is likely to design an economic
mechanism with high economic efficiency, but with little concern for
traditional systems metrics of computational efficiency, reliability,
security, complexity, and ease-of-use. Pure systems designers have
generally done the inverse. This is a direct consequence of a strict
interpretation of the policy/mechanism separation principle. Instead,
we advocate that systems designers embrace mechanism design as a
first-order concern to eventually produce systems than can be both
economically and computationally efficient.